\documentclass{article}
\usepackage{a4wide,epsfig,cite}
\usepackage{hyperref}

\def\sstyle{\scriptstyle}

\def\binom#1#2{\left(\begin{array}{@{}c@{}} #1\\#2\end{array}\right)}
\def\ket#1{|#1\rangle}
\def\n{\sstyle 0} 
\def\be{\begin{equation}}
\def\ee{\end{equation}}
\def\bea{\begin{eqnarray}}
\def\eea{\end{eqnarray}}
\begin{document}

\renewcommand{\thefootnote}{\arabic{footnote}}
\title{Refined Razumov-Stroganov conjectures for open boundaries}
\author{Jan de Gier$^1$ and Vladimir Rittenberg$^2$\\[5mm] {\small\it
    $^1$Department of Mathematics and Statistics, The University of  
  Melbourne, VIC 3010, Australia}\\
{\small\it$^2$Physikalisches Institut, Universit\"at Bonn, Nussallee 12, Bonn, Germany}} 

\date{\today}
\maketitle
\footnotetext[1]{\href{mailto:degier@ms.unimelb.edu.au}{{\tt
      degier@ms.unimelb.edu.au}}}
\footnotetext[2]{\href{mailto:vladimir@th.physik.uni-bonn.de}{{\tt vladimir@th.physik.uni-bonn.de }}}

\begin{abstract}
Recently it has been conjectured that the ground-state of a Markovian
Hamiltonian, with one boundary operator, acting in a link pattern
space is related to vertically and horizontally symmetric
alternating-sign matrices (equivalently fully-packed loop configurations
(FPL) on a grid with special boundaries). 
%The coefficient of the boundary
%operator which gives the boundary rates, was taken equal to one. 
We extend this conjecture by introducing an arbitrary boundary 
parameter. We show that the parameter dependent ground state
is related to refined vertically symmetric alternating-sign matrices i.e.
with prescribed configurations (respectively, prescribed FPL
configurations) in the next to central row.

We also conjecture a relation between the ground-state of a Markovian
Hamiltonian with two boundary operators and arbitrary coefficients and some
doubly refined (dependence on two parameters) FPL configurations. Our conjectures
might be useful in the study of ground-states of the O(1) and XXZ
models, as well as the stationary states of Raise and Peel models. 
\end{abstract}

\section{Introduction}
In a remarkable paper, Razumov and Stroganov (R-S) \cite{RazuS00} have looked at the
ground-state wave function of the ferromagnetic one-dimensional XXZ spin
1/2 chain with the asymmetry parameter $\Delta=-1/2$, odd number of sites
and periodic boundary condition. They noticed that for a small number of
lattice sites $L=2n+1$ (this was all numerics), the largest component and
the normalization are related to the number of $n\times n$ alternating-sign
matrices (ASMs) \cite{Bress99}. They conjectured that these coincidences are valid
for any number of sites. This conjecture, though not yet proven, triggered
a lot of other conjectures which brought together the study of quantum
chains and combinatorics.

It was realized in \cite{BatchGN01} that similar conjectures can be made in the O(1)
loop model with various boundary conditions. This led R-S \cite{RazuS01}, using the
bijection between ASMs and fully packed loops (FPLs) on a grid with
special boundary conditions, to a new conjecture (called hereafter
the R-S conjecture) which gives 
a much deeper connection between the ground-state wave function and ASMs.
Conjectures similar to the one of R-S were also made
for Hamiltonians with different boundary conditions \cite{RazuS01b,PearceRGN02}
relating them to different symmetry classes of ASMs.

In another development (relevant to the present paper), it was
understood that the results obtained in the O(1) model can be derived
writing the Hamiltonians in terms of generators of the Temperley-Lieb algebra
at the semi-group point \cite{PearceRGN02}, acting on the link pattern (or
equivalently the restricted solid on solid path) basis. Each Hamiltonian gives the
Markovian time evolution of a fluctuating interface. The class of
stochastic models of this type are called Raise and Peel (RP) models
(see \cite{GierNPR03} for physical properties and \cite{Pyatov04} for
several examples). Besides their interesting properties, these models
belong to a new universality class for non-equilibrium phenomena,
since the finite-size scaling spectrum of these models are given by conformal
field theory. 
 
In a very recent new development, Di Francesco \cite{difran04} proposed a refined
R-S conjecture. If one considers $A_n(j)$ the known number \cite{Zeilb96}
of $n\times n$ ASMs with a $1$ on top of the $j$th column one can define a generating
function $\psi_n(t)$.  According to Di Francesco's conjecture this
function coincides with the normalization of the ground-state wavefunction
of the $t$ dependent monodromy matrix. As for the other cases when
R-S type conjectures were made, not only the normalization,
but each component of the $t$-dependent ground-state gives information on
properties of ASMs.

In this paper we will present two new refined R-S type conjectures.
In Section \ref{se:1b} we present the first conjecture. We define a
Hamiltonian in terms of the $L$ generators of the boundary Temperley-Lieb
algebra. This Hamiltonian acts in the space of $2^L$ link patterns and
depends on a free parameter $a$. If this parameter is nonnegative, the
Hamiltonian gives the time evolution of a fluctuating interface. The
parameter $a$ controls the 
boundary rates. We have studied the ground-state wave function, which
gives the probability distribution function of the stationary state of the
stochastic process, as a function of $a$.
 
For $a=1$, it was observed in \cite{MitraNGB04}, a la Razumov and
Stroganov, that the ground-state wave function $\psi_0(1)$ for a
system of size $L$ is related to vertically and horizontally symmetric
alternating sign matrices (VHASMs) of size $2L+3$. In the present
paper we study $\psi_0(a)$ and make a refined R-S conjecture. As an
application, we give the average number of $\pm 1$'s on the $(L+1)$-st row
of a VHASM matrix of size $2L+3$. Other results can also be used to
study the properties of stationary states of the stochastic process.

Although, it is not the main subject of this paper, we also comment
on the connection of the Hamiltonian with the XXZ quantum chain with
diagonal boundary conditions. We furthermore briefly discuss the properties of
the spectrum of the Hamiltonian in the continuum limit.

The second conjecture is presented in Section \ref{se:2b}. We add to the
Hamiltonian considered in Section \ref{se:1b} a second boundary operator with a
coefficient $b$. The new Hamiltonian depends therefore on two parameters.
The space of link patterns in which the new Hamitonian acts is
different from the one in which the Hamiltonian with only one generator
acts. Remarkably, the dimensions of the subspaces in which the two
ground-states have components are the same. 

We conjecture that the ground-state $\psi_0(a,b)$ of the Hamiltonian with
$L-1$ bulk and $2$ boundary generators is related to weighted FPL diagrams on a grid of
dimension $(L+1) \times L$ with special boundary conditions. A given
configuration gets a factor $a$ for each vertical segment at the boundary and a factor $b$ for
each horizontal segment at the boundary.
% (the two kind of segments are defined in the text).

\section{The refined one-boundary R-S conjecture}
\label{se:1b}
The Hamiltonian we wish to study is given by
\be
H = a(1-f_-) + \sum_{j=1}^{L-1} (1-e_j), 
\label{eq:ham}
\ee
where the generators $f_-$ and $e_i$ satisfy the one-boundary Temperley-Lieb
algebra, otherwise known as the blob algebra \cite{MartS94},
\bea
e_i^2 &=& e_i,\nonumber\\
e_ie_{i\pm 1}e_i &=& e_i,\nonumber\\
e_ie_j &=& e_je_i\quad {\rm for}\quad |i-j| \geq 2,
\label{eq:TLdef}\\ 
f_-^2 &=& f_-,\nonumber\\
e_1 f_-e_1 &=& e_1.\nonumber
\eea
The Hamiltonian given by (\ref{eq:ham}) describes the time evolution
of a stochastic process of a fluctuating interface with a boundary, see
\cite{Pyatov04} for the case $a=1$. Here we are interested in the
stationary state only. 

\begin{figure}[h]
\centerline{
\begin{picture}(150,20)
\put(0,0){\epsfxsize=150pt\epsfbox{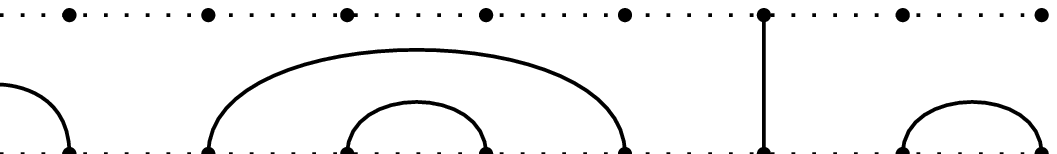}}
\end{picture}}
\caption{An $8\times 2$ strip with link pattern $)(())|()$.}
\label{fig:strip}
\end{figure}

The Temperley-Lieb algebra has a well known loop representation in the
space of link patterns \cite{TempL71,Mart91}. Let us
briefly recall this representation. Consider an $L\times 2$ strip whose
$L$ bottom sites may be connected to each other or to the site
directly above it by non-crossing arcs. In the case of the
one-boundary Temperley-Lieb algebra, sites can also be connected to
the left boundary of the strip. See Fig.~\ref{fig:strip} for an
example of an $8\times 2$ strip. We call a particular way in
which loops are connected a {\it link pattern} or  {\it connectivity},
and denote the linear span of link patterns on $L$ sites by LP$_L$. 

On a link pattern $\pi\in {\rm LP}_L$ the generator $e_i$ acts in
the following way: If site $i$ is connected to $k$ and site $i+1$ to $l$,
$e_i$ connects $i$ with $i+1$ and $k$ with $l$. If $i$ is connected to
$k$ and $i+1$ to the top row (or left boundary) of the strip, then $k$
gets connected to the top row (or left boundary) and $i$ to $i+1$;
similarly, if $i$ is connected to the top row (or left boundary) and $i+1$ is
connected to $k$. Lastly, if both $i$ and $i+1$ are connected to the
top row (or left boundary), they get connected to each other. If site $1$ is
connected to $i$, the generator $f_-$ acts by connecting both site $1$
and site $i$ to the left boundary of the strip. If site $1$ is
connected to the top row it gets connected to the left boundary, while
if site $1$ were connected to the left boundary, $f_-$ acts as the identity.

Link patterns can be described using a parentheses notation: If site
$i$ is connected to site $j$ we put an opening parenthesis ``(`` at
$i$ and a closing parenthesis ``)'' at $j$. If site $i$ is connected to
the left boundary we put a closing parenthesis at $i$. Sites that are
connected to the top of the strip are denoted by vertical bars. The
link patterns for $L=2$ are thus given by   
\be
||,\quad )|,\quad )),\quad (),
\ee
which respectively mean that (i) the two sites are connected to the
top of the strip, (ii) the first is connected to the left boundary
while the second is connected to the top, (iii) both are connected to
the left boundary and (iv) the two sites are connected to each
other. The dimension of the space of link patterns for the
one-boundary Temperley-Lieb algebra on $L$ sites is 
\be
\dim {\rm LP}_L = 2^L.
\ee
Due to the semi-group structure %\cite{GierNPR} 
of the algebra (\ref{eq:TLdef}), the Hamiltonian (\ref{eq:ham}) has a
positive spectrum and a unique 
ground-state energy $E_0=0$ in ${\rm LP}_L$. We will be
interested in the corresponding eigenvector $\psi_0$ as a function of
the parameter $a$,
\be
H\psi_0(a) =0. 
\ee
This eigenvector lies in the subspace LP$_L^0$ spanned by the link
patterns without vertical bars. Its dimension is equal to
\be
\dim {\rm LP}_L^0 = \binom{L}{\lfloor L/2\rfloor}.
\ee

\subsection{Fully packed loops}
\begin{figure}[h]
\centerline{
\begin{picture}(225,90)
\put(0,10){\epsfxsize=150pt\epsfbox{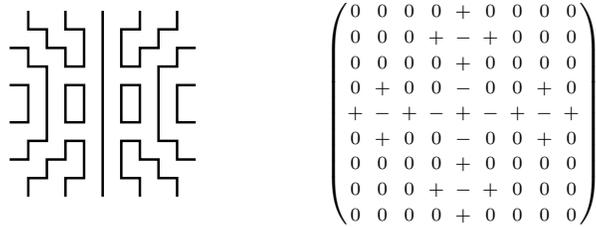}}
\put(120,40){$
\setlength{\arraycolsep}{2pt}
\renewcommand{\arraystretch}{0.8}
\left(\begin{array}{@{}*{9}c@{}}
\n & \n & \n & \n & \sstyle + & \n & \n & \n & \n\\
\n & \n & \n & \sstyle + & \sstyle - & \sstyle + & \n & \n & \n\\
\n & \n & \n & \n & \sstyle + & \n & \n & \n & \n\\
\n & \sstyle + & \n & \n &\sstyle  - & \n & \n & \sstyle + & \n\\
\sstyle + & \sstyle - & \sstyle + & \sstyle - & \sstyle + & \sstyle
- & \sstyle + & \sstyle - & \sstyle +\\ 
\n & \sstyle + & \n & \n & \sstyle - & \n & \n & \sstyle + & \n\\
\n & \n & \n & \n & \sstyle + & \n & \n & \n & \n\\
\n & \n & \n & \sstyle + & \sstyle - & \sstyle + & \n & \n & \n\\
\n & \n & \n & \n & \sstyle + & \n & \n & \n & \n
\end{array}\right)
\renewcommand{\arraystretch}{1}
$}
\end{picture}}
\caption{A vertically and horizontally symmetric FPL diagram on a grid
  of size $9$ and the corresponding alternating-sign matrix.}
\label{fig:grid}
\end{figure}

In \cite{Pyatov04,MitraNGB04} the Hamiltonian (\ref{eq:ham}) was studied
for $a=1$ and it was observed that $\psi_0(1)$ for size $L$ is related to
vertically and horizontally symmetric alternating-sign matrices
(VHASMs). In particular it was conjectured that the sum of the
components of $\psi_0(1)$ for size $L$ is equal to $A_{\rm VH}(2L+3)$,
the total number of VHASMs of size $2L+3$. This number is known
\cite{Kupe96} and given by
\bea
A_{\rm VH}(4n\pm1) &=& A_{\rm V}(2n\pm 1)N_8(2n) ,\\
A_{\rm V}(2n+1) &=& \prod_{k=0}^{n-1} (3k+2)
\frac{(6k+3)!(2k+1)!}{(4k+2)!(4k+3)!} = 1,3,26,646,\ldots,\\
N_{8}(2n) &=& \prod_{k=0}^{n-1} (3k+1)
\frac{(6k)!(2k)!}{(4k)!(4k+1)!}=1,2,11,170,\ldots,
\eea
where $A_{\rm V}(2n+1)$ is the number of vertically symmetric
$(2n+1)\times (2n+1)$ alternating-sign matrices, and $N_8(2n)$ the
number of cyclically symmetric transpose complement plane partitions
in a box of size $2n\times 2n\times 2n$. By a well known bijection
\cite{BatchBNY96}, $A_{\rm VH}(2L+3)$ is also equal to
the total number of vertically and horizontally symmetric fully packed
loop (FPL) diagrams on grids of size $2L+3$. An example of a
vertically and horizontally symmetric FPL diagram on a grid of size
$9$ is given in Fig.~\ref{fig:grid}. For later purposes we note that
the nonzero entries in the ASM correspond to straight loop segments of
length $2$.

\begin{figure}[h]
\centerline{
\begin{picture}(150,75)
\put(0,0){\epsfxsize=150pt\epsfbox{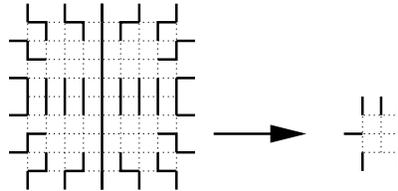}}
\end{picture}}
\caption{Reduction of a $2L+3$ grid with vertical and horizontal
  symmetry to an $L \times L$ patch. }
\label{fig:patch}
\end{figure}

\begin{figure}[h]
\centerline{
\begin{picture}(100,85)
\put(0,0){\epsfxsize=100pt\epsfbox{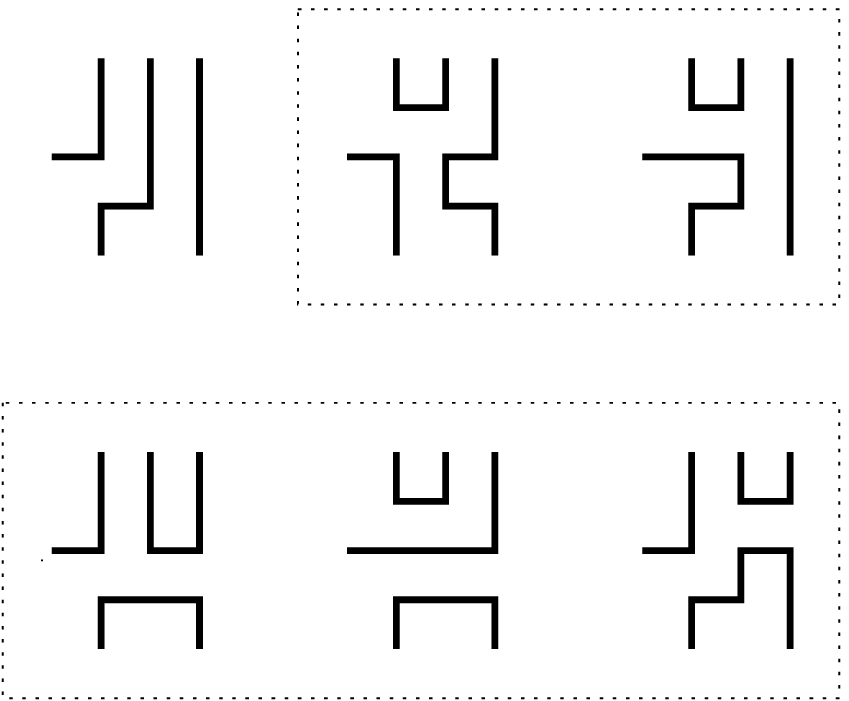}}
\put(-2,63){$\sstyle 1$}
\put(10,45){$\sstyle 2$}
\put(22,45){$\sstyle 3$}
\end{picture}}
\caption{FPL diagrams for $L=3$. Diagrams with the same link
  pattern are grouped together. The last FPL diagram corresponds to
  the VHASM given in Fig.~\ref{fig:grid}.}
\label{fig:gsL3}
\end{figure}

The vertical and horizontal symmetry constraint enforces certain edges
to contain loop segments, and also that the four quadrants of the grid
are mirror images of each other. We therefore only have to consider
FPL diagrams on an $L\times L$ patch, see Fig.~\ref{fig:patch}.
The six FPL patterns that can be drawn on the $3\times3$ patch are depicted
in Fig.~\ref{fig:gsL3}. For each FPL diagram one can define a
link pattern describing the way the external edges are connected to
each other by the loop segments: We number from $1$ to 
$L$ the external edges that contain loop segments on the left and
bottom of the $L\times L$ patch, as in the top left diagram of
Fig.~\ref{fig:gsL3}. If edge $i$ is connected to edge $j$ we 
put an opening parenthesis ``(`` at $i$ and a closing parenthesis
``)'' at $j$. If edge $i$ is connected to one of the external edges on
the top we put a closing parenthesis at $i$. The link patterns in
Fig.~\ref{fig:gsL3} are thus ))), ()) and )() and diagrams with the same
link pattern are grouped together. 

Following Razumov and Stroganov \cite{RazuS01} it was further
conjectured that not only the sum of components of $\psi_0(1)$ is
related to a counting problem, but also each coefficient of
$\psi_0(1)$. Writing
\be
\psi_0(1) = \sum_{\pi \in {\rm LP}_L^0} c_L(\pi) \ket\pi,
\ee
the conjecture states that $c_L(\pi)$ is equal to the number of FPL
diagrams on the $L\times L$ patch with link pattern $\pi$. For example
for $L=3$ we obtain 
\be 
\psi_0(1) = (1,2,3)
\ee
on the basis $\{))),()),)()\}$, indeed corresponding to the
enumeration of FPL diagrams with the same link pattern, see
Fig.~\ref{fig:gsL3}.

\subsection{A refined conjecture}
For general values of the parameter $a$ we find that the conjecture above
is refined in the following way. If we write
\be
\psi_0(a) = \sum_{\pi \in {\rm LP}_L^0} c_L(\pi,a) \ket\pi,
\ee
the coefficient $c_L(\pi,a)$ will be a polynomial in $a$. We have
observed from exact calculations for small values of $L$ that the
coefficient of $a^j$ in $c_{2n}(\pi,a)$ (resp. $c_{2n+1}(\pi,a)$) enumerates
FPL diagrams on the $L\times L$ patch with connectivity $\pi$ {\it
  and} having $2j$ (resp. $2j+1$) vertical line segments of length $2$
measured from the top. We will illustrate this by two examples. 

\begin{figure}[h]
\centerline{
\begin{picture}(100,85)
\put(0,0){\epsfxsize=100pt\epsfbox{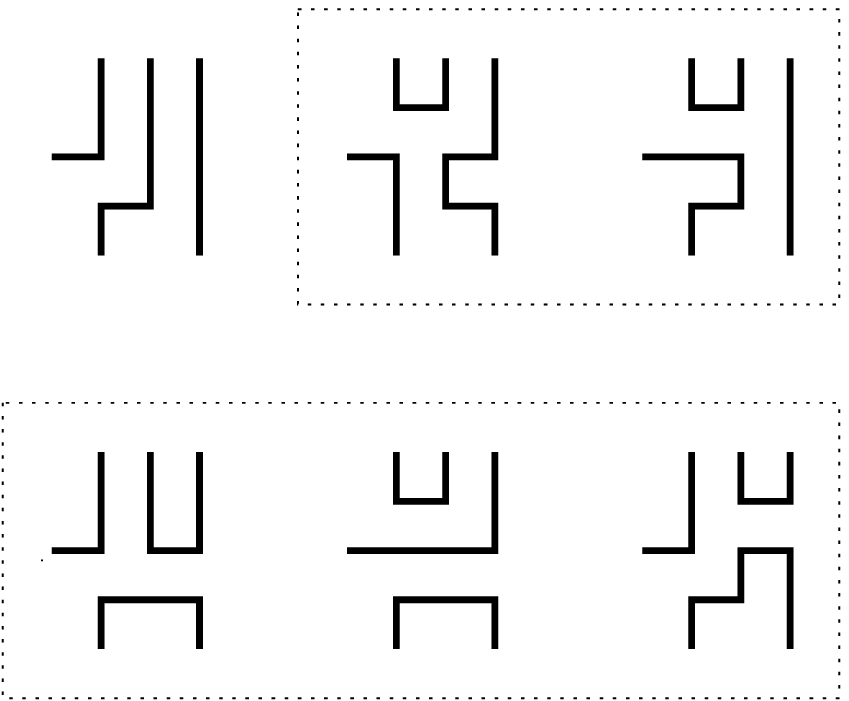}}
\put(3,74){$\sstyle a$}\put(38,75){$\sstyle 1$}\put(72,75){$\sstyle 1$}
\put(3,27){$\sstyle a$}\put(38,27){$\sstyle 1$}\put(72,27){$\sstyle 1$}
\end{picture}}
\caption{Weighted FPL diagrams for $L=3$. A diagram with $2j+1$ vertical line
  segments in the top row is assigned a weight
  $a^j$. Diagrams with the same link pattern are grouped together.}
\label{fig:gsL3W}
\end{figure}

For $L=3$ we find
\be
\psi_0(a) = (a,2,2+a),
\ee
on the basis $\{))),()),)()\}$. Assign to each FPL diagram on the
$3\times 3$ patch a weight $a^j$ if 
it has $2j+1$ vertical line segments on the first row, as in
Fig.~\ref{fig:gsL3W}. The coefficient $c_3(\pi,a)$ of $\psi_0(a)$ is
then given by the sum of weighted FPL diagrams having link pattern $\pi$.  

\begin{figure}[h]
\centerline{
\begin{picture}(320,300)
\put(0,0){\epsfxsize=320pt\epsfbox{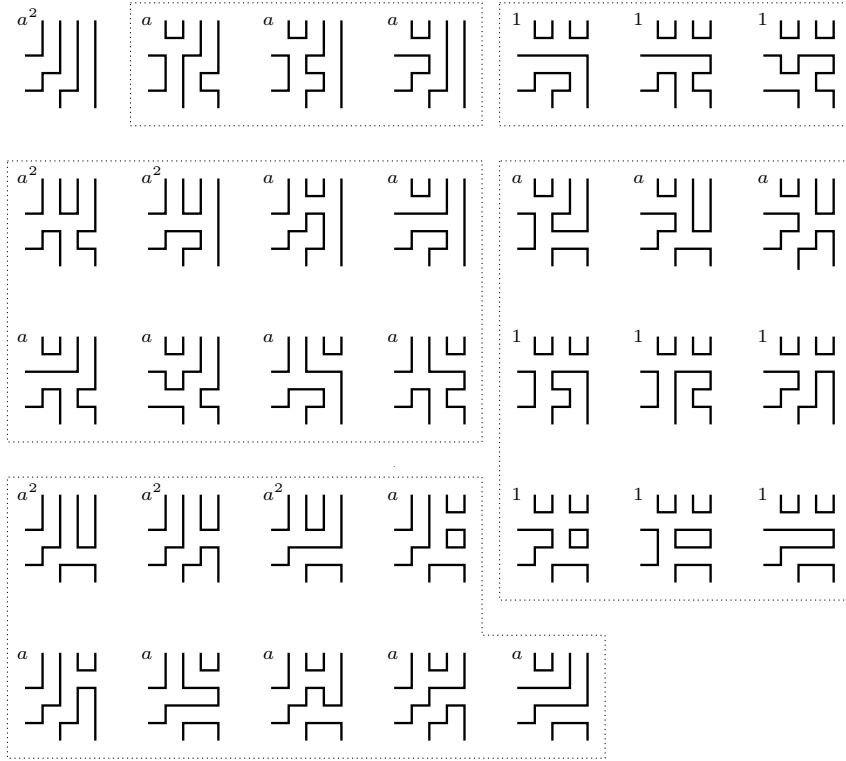}}
\put(4,278){$\sstyle a^2$}\put(51,278){$\sstyle a$}\put(97,278){$\sstyle a$}\put(144,278){$\sstyle a$}\put(191,278){$\sstyle 1$}\put(237,278){$\sstyle 1$}\put(284,278){$\sstyle 1$}
\put(4,218){$\sstyle a^2$}\put(51,218){$\sstyle a^2$}\put(97,218){$\sstyle a$}\put(144,218){$\sstyle a$}\put(191,218){$\sstyle a$}\put(237,218){$\sstyle a$}\put(284,218){$\sstyle a$}
\put(4,158){$\sstyle a$}\put(51,158){$\sstyle a$}\put(97,158){$\sstyle a$}\put(144,158){$\sstyle a$}\put(191,158){$\sstyle 1$}\put(237,158){$\sstyle 1$}\put(284,158){$\sstyle 1$}
\put(4,98){$\sstyle a^2$}\put(51,98){$\sstyle a^2$}\put(97,98){$\sstyle a^2$}\put(144,98){$\sstyle a$}\put(191,98){$\sstyle 1$}\put(237,98){$\sstyle 1$}\put(284,98){$\sstyle 1$}
\put(4,38){$\sstyle a$}\put(51,38){$\sstyle a$}\put(97,38){$\sstyle a$}\put(144,38){$\sstyle a$}\put(191,38){$\sstyle a$}
\end{picture}}
\caption{Weighted FPL diagrams for $L=4$. A diagram with $2j$ vertical
  line segments in the top row is assigned a weight
  $a^j$. Diagrams with the same link pattern are grouped together.}
\label{fig:gsL4W}
\end{figure}

For the second example we look at the even system $L=4$ and obtain
\be
\psi_0(a)=(a^2,3a(2+a),2a(3+a),3a,3(2+a),3),
\ee
on the basis $\{)))), ))(), )()), ())), ()(), (())\}$. These
components are indeed recovered by the enumeration of weighted FPL diagrams,
if we assign to each FPL diagram a weight $a^j$ if it has $2j$
vertical line segments in the top row, as in Fig.~\ref{fig:gsL4W}.
The refined conjecture formulated above is similar to but differs from
Di Francesco's recent results \cite{difran04}. 

\subsection{An application}
The conjectures above can be used to derive properties of FPL diagrams
orASMs. For example, the normalization of the one-boundary ground-state
\be
Z^{(1)}(a) = \sum_{\pi\in{\rm LP}^0_L} c_L(\pi,a),
\ee
is conjectured to be equal to the partition function of $(2L+3)\times
(2L+3)$ VHASMs in which each $-1$ on the row below the horiontal
symmetry axis is given a weight $a$. The average density $\rho_L(a)$
of $-1$'s on this row is thus given by 
\be   
\rho(a) = \frac1L \frac{{\rm d} \log Z^{(1)}(a)} {{\rm d}a}.
\ee
While we have not found a general expression for $Z^{(1)}(a)$, we
have found numerically that 
\be
\rho_L(1) =
\left\{ 
\renewcommand{\arraystretch}{2.3}
\begin{array}{@{}cc@{}}
\displaystyle
\frac{3L+8}{8(2L+3)}\quad {\rm for\;even\;}L \\
\displaystyle
\frac{3(L^2-1)}{8L(2L+3)}\quad {\rm for\;odd\;}L
\end{array}\right.
\ee
In the thermodynamic limit $\rho_\infty(1)=3/8$ whereas one would
expect a value of $1/3$ in a completely random region. 

\subsection{Relation to the XXZ chain with diagonal
  boundary conditions}
We would like to comment about the spectrum of the Hamiltonian given in
(\ref{eq:ham}). It is convenient to parametrize $a$ defined in
(\ref{eq:ham}) as follows:
\be
a=\frac{3}{1+2\cos\delta} > 0.
\ee
The spectrum of $H$ acting in the link pattern space coincides with that
of the Hamiltonian of the XXZ spin 1/2 quantum chain with
diagonal boundary conditions and asymmetry parameter $\Delta = -1/2$ \cite{Nichols}: 
\bea
H_{\rm D} &=& -\frac12 \left\{ \sum_{j=1}^{L-1}
\left(\sigma_j^x\sigma_{j+1}^{\rm x} + \sigma_j^y\sigma_{j+1}^y -
\frac12 \sigma_j^z\sigma_{j+1}^z \right)\right. \nonumber \\
&&\left. {} + \frac{\sqrt{3}}{2} \left(
\tan\left(\frac{\pi}{6}+\frac{\delta}{2}\right) \left(\sigma_1^z-1\right) +
\tan\left(\frac{\pi}{6}-\frac{\delta}{2} \right) \left(\sigma_L^z-1\right)
\right)\right\} + \frac34 (L-1), 
\eea
where $\sigma^x$, $\sigma^y$ and $\sigma^z$ are Pauli matrices. For
$a\geq 1$ ($0\leq\delta< 2\pi/3$) the Hamiltonian $H_{\rm D}$ is
Hermitian. Notice that although
in (\ref{eq:ham}) the parameter $a$ appears in one boundary generator only, it
appears in both boundary terms in the expression for $H_{\rm D}$.
Moreover there exists a similarity transformation which relates $H$ to
$H_{\rm D}$. This last observation is not trivial since Jordan cell structures
often appear in stochastic processes and two Hamiltonians might have the
same spectrum but might not be related by a similarity transformation
\cite{Nichols}.

Since $H$ describes a stochastic process it has a positive spectrum with a
unique ground-state of zero energy. The same properties are therefore
inherited by $H_{\rm D}$. In the special case $a=1$ ($\delta=0$), using different
methods, in \cite{FendSN03} it was shown that $H_{\rm D}$ has a positive spectrum and
a zero ground-state energy for all system sizes.

We will now shortly comment on the spectrum of $H_{\rm D}$ (or $H$) in the
continuum limit.
%For the domain $a\geq 1$, the finite-size scaling spectrum of $H_{\rm
%  D}$ is known. 
It
is independent of $a$ \cite{AlcBB87}, it is the same for $L$ even and odd
and is given by the Gauss model. A compact way to describe the spectrum is to
use \cite{GierNPR} the sum of the characters of the $[1/24]$ and $[3/8]$
representations of the $c=1$ Ramond representations of $N=2$ superconformal field
theory \cite{EguY90}. 
%We didn't study the finite-size scaling spectrum for $a<1$. All what
%is known is that 
For $a=0$ the Hamiltonian $H_{\rm D}$ is $U_q(sl(2))$-invariant and
the finite-size scaling spectrum is different \cite{BaueS89}.

\section{The two-boundary case}
\label{se:2b}
We can extend the Hamiltonian (\ref{eq:ham}) to also include a
boundary term at site $L$,
\be
H = a(1-f_-) + b(1-f_+) + \sum_{j=1}^{L-1} (1-e_j).
\label{eq:ham2}
\ee
As before, the generators $f_-$ and $e_i$ satisfy the relations of the
one-boundary Temperley-Lieb algebra, but with the additional relations
\bea
f_+^2 &=& f_+,\nonumber\\
e_{L-1} f_+e_{L-1} &=& e_{L-1},\\
IJI &=& I,\qquad JIJ\;=\;J, \nonumber
\eea
where
\bea
I=\prod_{i=0}^{L/2-1} e_{2i+1},\qquad J=f_- \prod_{i=1}^{L/2-1} e_{2i}
f_+,\qquad {\rm for\; even\;} L,\\
I=f_-\prod_{i=1}^{(L-1)/2} e_{2i},\qquad J=\prod_{i=1}^{(L-1)/2} e_{2i-1}
f_+,\qquad {\rm for\; odd\;} L.
\eea
In the link pattern representation, see Fig.~\ref{fig:strip}, loops
are now also allowed to be connected to the right boundary of the
strip. Connections to the right boundary are facilitated by $f_+$ in
much the same way as they are for the left boundary by $f_-$. 

We wish to study the ground-state of the Hamiltonian (\ref{eq:ham2}), which
lies in the subspace of link patterns for which all sites are connected.
One may furthermore choose not to make a distinction between the left
and right boundary, but to identify them, and this is what we will do
in the following. As before we denote link patterns by sequences
of parentheses and vertical bars. Parentheses are used for sites
that are connected to each other while a vertical bar will now stand
for sites connected to the boundary. For example, the ground-state
sector of the Hamiltonian (\ref{eq:ham2}) with identified boundaries
for $L=4$ is given by the linear span of the six link patterns
\be
||||,\quad ()||,\quad |()|,\quad ||(),\quad ()(),\quad (()).
\ee
We will denote the linear span of such link patterns by LP$_L^*$. Its
dimension is given by 
\be
\dim {\rm LP}_L^* = \binom{L}{\lfloor L/2\rfloor},
\ee
which is the same as $\dim {\rm LP}_L^0$.

\begin{figure}[h]
\centerline{
\begin{picture}(70,60)
\put(0,0){\epsfxsize=70pt\epsfbox{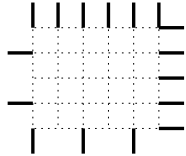}}
\end{picture}}
\caption{A $6\times 5$ patch corresponding to the ground-state of the
  Hamiltonian (\ref{eq:ham2}) for $L=5$ and identified boundaries.}
\label{fig:2b-odd-patch}
\end{figure}

\begin{figure}[h]
\centerline{
\begin{picture}(150,150)
\put(0,0){\epsfxsize=150pt\epsfbox{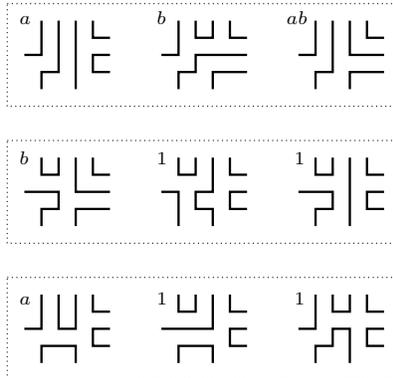}}
\put(5,135){$\sstyle a$}\put(57,135){$\sstyle b$}\put(106,135){$\sstyle ab$}
\put(5,82){$\sstyle b$}\put(57,82){$\sstyle 1$}\put(109,82){$\sstyle
  1$}
\put(5,29){$\sstyle a$}\put(57,29){$\sstyle 1$}\put(109,29){$\sstyle 1$}
\end{picture}}
\caption{Weighted FPL diagrams for $L=3$. A diagram with $2j+1$ vertical line
  segments in the top row and $2k$ horizontal line segments in the
  rightmost column is assigned a weight $a^jb^k$. Diagrams with the same
  link pattern are grouped together.} 
\label{fig:2bgsL3W}
\end{figure}

%\subsection{A doubly refined conjecture}
We will now formulate a doubly refined conjecture based on
observations for small systems. For our purposes we first need to formulate
a new observation:  For $a=b=1$ and {\it odd} $L$ the ground-state is
obtained by counting FPL configurations with appropriate link
patterns on an $(L+1)\times L$ patch with boundary conditions as in
Fig.~\ref{fig:2b-odd-patch}.\footnote{In \cite{MitraNGB04} it was observed that
the ground-state is also obtained by counting FPL configurations with
appropriate link patterns on an $L\times(L+1)/2$ patch. Compared to
this observation, ours gives rise to an overall factor (equal to
$A_{\rm V}(L+2)$) in each of the components. The relation between both
observations is not clear.}  
We can now formulate a doubly refined conjecture for odd
systems. Writing  
\be
\psi_0(a,b) = \sum_{\pi \in {\rm LP}_L^*} c_L(\pi,a,b) \ket\pi,
\ee
the coefficient $c_L(\pi,a,b)$ is a polynomial in $a$ and $b$. From
exact calculations for small values of $L=2n+1$ we observed that the 
coefficient of $a^jb^k$ in $c_{2n+1}(\pi,a,b)$ enumerates FPL diagrams
on the $(L+1)\times L$ patch with connectivity $\pi$ {\it and} having
$2j+1$ vertical line segments in the top row and $2k$
horizontal line segments in the last column. For $L=3$ we find for example
\be
\psi_0(a,b) = (a+b+ab,2+b,2+a),
\ee
on the basis $\{|||,()|,|()\}$, indeed corresponding to the weighted
enumeration of FPL diagrams, see Fig.~\ref{fig:2bgsL3W}. In addition we note that the normalization for odd system sizes,
\be
Z_{2n+1}^{(2)}(a,b) = \sum_{\pi \in {\rm LP}^*_{2n+1}} c_{2n+1}(\pi,a,b),
\ee
factorises completely, 

\be
Z_{2n+1}^{(2)}(a,b) = \tilde{Z}^{(1)}_{2n+1}(a) \tilde{Z}^{(1)}_{2n+1}(b).
\ee

\begin{figure}[h]
\centerline{
\begin{picture}(80,70)
\put(0,0){\epsfxsize=80pt\epsfbox{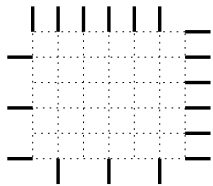}}
\end{picture}}
\caption{A $7\times 6$ patch corresponding to the ground-state of the
  Hamiltonian (\ref{eq:ham2}) for $L=6$ and identified boundaries.}
\label{fig:2b-even-patch}
\end{figure}

For $a=b=1$ and {\it even} $L$ it was conjectured in
\cite{MitraNGB04} that the 
coefficients of the ground-state enumerate FPL configurations with
appropriate link patterns on an $(L+1)\times L$ patch with boundary
conditions as in Fig.~\ref{fig:2b-even-patch}. 
The doubly refined conjecture for even systems is formulated as follows. Write the
ground-state as
\be
\psi_0(a,b) = \sum_{\pi \in {\rm LP}_L^*} c_L(\pi,a,b) \ket\pi,
\ee
then the coefficient $c_L(\pi,a,b)$ is a polynomial in $a$ and $b$. We have
observed from exact calculations for small values of $L=2n$ that the
coefficient of $a^jb^k$ in $c_{2n}(\pi,a,b)$ enumerates FPL diagrams
on the $(L+1)\times L$ patch with connectivity $\pi$ {\it and} having
$2j$ or $2j+1$ vertical line segments in the top row and $2k$
horizontal line segment in the last column.

\section{Conclusion}
This paper contains two conjectures concerning the relation between
parameters dependent ground-states of Hamiltonians describing stochastic
processes and combinatorial properties of FPLs on grids with special boundary
conditions.

We hope that the existence of more conjectures of this kind will help to
bring, finally, a proof of the many known parameter-fixed  conjectures
which are already in the literature (see the references mentioned in the
introduction).

In both conjectures the parameters are related to boundary operators
and the bookkeeping of the FPLs is also done by looking at how the loops
touch the boundaries. This is in contrast to the refined conjecture of Di
Francesco \cite{difran04} (which was a source of inspiration for the present paper) in
which the parameter changes the bulk interaction but the bookkeeping of
the FPLs still counts the way the loops touch the boundary. The second
difference between the two approaches can be understood in the following
way. If, in analogy with statistical physics, one interprets the parameters
dependent ground-state as a partition function depending on
fugacities, see e.g. \cite{BrakGR04} for a justification of this
interpretation, our two conjectures give informations on the number of
``defects" on the boundary (for example the number of non-zero entries
in the VHASM matrices on a row). The refined conjecture of
\cite{difran04} gives the space distribution of one ``defect" (the
position of $1$ in the first row of ASMs). 

\section{Acknowledgement}
We would like to thank A. Nichols for discussions. This work was done
within the European Commision network HPRN-CT-2002-00325. Financial
support from the Deutsche Forschunggemeinshaft and the Australian Research
Council are gratefully achnowledged.


\begin{thebibliography}{99}
\bibitem{RazuS00} Yu.G. Stroganov, 2001 {\it The importance of being odd},
  J. Phys. A {\bf 34} L179--L185; A.V. Razumov and
  Yu.G. Stroganov, 2001 {\it Spin chains   and combinatorics}, J. Phys. A
  {\bf 34} 3185--3190. 

\bibitem{Bress99} D. Bressoud, 1999 {\it Proofs and confirmations. The
  story of the alternating sign matrix conjecture}, Cambridge
  University Press.

\bibitem{BatchGN01} M.T. Batchelor, J. de Gier and B. Nienhuis, 2001 {\it The
  quantum symmetric XXZ chain at $\Delta=-1/2$, alternating sign
  matrices and plane partitions}, J. Phys. A {\bf 34} L265--L270.  

\bibitem{RazuS01} A.V. Razumov and Yu.G. Stroganov, 2004 {\it Combinatorial
  nature of gound state vector of O(1) loop model},
  Theor. Math. Phys. {\bf 138} 333--337,
  \href{http://arxiv.org/math.CO/0104216}{{\tt math.CO/0104216}} 

\bibitem{RazuS01b} A.V. Razumov and Yu.G. Stroganov, 2001 {\it Spin chains
  and combinatorics: twisted boundary conditions}, J. Phys. A {\bf 34}
  5335--5340, \href{http://arxiv.org/cond-mat/0102247}{{\tt
  cond-mat/0102247}}; A.V. Razumov and Yu.G. Stroganov, 2001 {\it O(1) loop 
  model with different boundary conditions and symmetry classes of
  alternating-sign matrices},
  \href{http://arxiv.org/cond-mat/0108103}{{\tt cond-mat/0108103}}. 

\bibitem{PearceRGN02}  P.A. Pearce, V. Rittenberg, J. de Gier and
  B. Nienhuis, 2002 {\it Temperley-Lieb stochastic processes}, J. Phys. A
  {\bf 35} L661-L668,
  \href{http://arxiv.org/math-ph/0209017}{\tt math-ph/0209017}.
 
\bibitem{GierNPR03} J. de Gier, B. Nienhuis, P.A. Pearce and
  V. Rittenberg, 2003 {\it Stochastic processes and conformal invariance},
  Phys. Rev. E {\bf 67} 016101--016104; 2004 {\it The raise and
  peel model of a fluctuating interface}, J. Stat. Phys. {\bf 114} 1--35.

\bibitem{Pyatov04} P. Pyatov, 2004 {\it Raise and peel models and Pascal's
  hexagon combinatorics}, \href{http://arxiv.org/math-ph/0406025}{{\tt
  math-ph/0406025}}.

\bibitem{difran04} P. Di Francesco, 2004 {\it A refined Razumov-Stroganov conjecture}, 
\href{http://arxiv.org/cond-mat/0407477}{{\tt cond-mat/0407477}}

\bibitem{Zeilb96} D. Zeilberger, 1996 {\it Proof of the refined alternating
  sign matrix conjecture}, New York J. Math. {\bf 2} 59--68

\bibitem{MitraNGB04} S. Mitra, B. Nienhuis, J. de Gier and
  M.T. Batchelor, 2004 {\it Exact expressions for correlations in the
  ground state of the dense O(1) loop model},
  \href{http://arxiv.org/cond-mat/0401245}{{\tt cond-mat/0401245}}

\bibitem{MartS94} P.P. Martin and H. Saleur, 1994 {\it The blob algebra
  and the periodic Temperley-Lieb algebra}, Lett. Math. Phys. {\bf 30}
  189--206; 

\bibitem{TempL71} H.N.V. Temperley and E.H. Lieb, 1971 {\it Relations
  between the `percolation' and `colouring' problem and other
  graph-theoretical problems associated with regular planar lattices:
  some exact results for the `percolation' problem}, Proc. R. Soc. A
  {\bf 322} 251--280

\bibitem{Mart91} P.P. Martin, 1991 {\it Potts models and related problems
  in statistical mechanics} (Singapore: World Scientific)

\bibitem{Kupe96} G. Kuperberg, 2002 {\it Symmetry classes of
  alternating-sign matrices under one roof}, Ann. Math. {\bf 156}
  835--866; \href{http://arxiv.org/math.CO/0008184}{\tt
  math.CO/0008184}

\bibitem{BatchBNY96} M.T. Batchelor, H.W.J. Bl\"ote, B. Nienhuis and
  C.M. Yung, 1996 {\it Critical behaviour of the fully packed loop
  model on the square lattice}, J. Phys. A {\bf 29} L399--L404;
  J. Propp, 2001 {\it The many faces of alternating-sign matrices},
  Discrete Mathematics and Theoretical Computer Science Proceedings AA
  43--58

%\bibitem{Sloane} N.J.A. Sloane, 2004, {\it The on-line encyclopedia
%  of integer sequences}, \url{http://www.research.att.com/~njas/sequences/}

\bibitem{Nichols} A. Nichols, private communication.

\bibitem{FendSN03} P. Fendley, K. Schoutens and B. Nienhuis, {\it
  Lattice fermion models with supersymmetry}, J. Phys. A {\bf 36}
  (2002), 12399--12424; X. Yang and P. Fendley, {\it Non-local space-time
  supersymmetry on the lattice}, \href{http://arxiv.org/cond-mat/0404682}{{\tt
  cond-mat/0404682}}. M. Beccaria and G.F. De Angelis, {\it Exact
  ground state and finite size scaling in a supersymmetric lattice
  model} \href{http://arxiv.org/cond-mat/0407752}{{\tt
  cond-mat/0407752}}.   

\bibitem{AlcBB87} F.C. Alcaraz, M.N. Barber and M.T. Batchelor, {\it
 Conformal invariance and the spectrum of the XXZ chain},
 Phys. Rev. Lett. {\bf 58} (1987), 771--774; F.C. Alcaraz, M. Baake, 
 U. Grimm and V. Rittenberg, {\it The modified XXZ Heisenberg chain,
 conformal invariance and the surface exponents of $c<1$ systems},
 J. Phys. A {\bf 22} (1989), L5--L12; U. Grimm and V. Rittenberg, {\it
 The modified XXZ Heisenberg chain, conformal invariance, surface
 exponents of $c<1$ systems, and hidden symmetries of the finite
 chains}, Int. J. Mod. Phys. B {\bf 4} (1990), 669--978.

\bibitem{GierNPR} J. de Gier, A. Nichols, P. Pyatov and V.Rittenberg,
  (to be published).

\bibitem{EguY90} F. Ravanini and S.-K. Yang, {\it $C$-disorder fields
  and $\Gamma(2)$-invariant partition functions in parafermionic
  conformal field theories}, Nucl. Phys. B
  {\bf 295} (1988), 262--276; T. Eguchi and S-K. Yang, {\it $N=2$ superconformal
  models as topological field theories}, Mod. Phys. Lett. {\bf 5} (1990)
  1693--1701; H. Saleur and N.P. Warner, {\it Lattice models and $N=2$
  supersymmetry}, \href{http://arxiv.org/hep-th/9311138}{{\tt
  hep-th/9311138}}, 49 pp.

\bibitem{BaueS89} M. Bauer and H. Saleur, {\it On some relations
  between local height probabilities and conformal invariance},
  Nucl. Phys. B {\bf 320} (1989), 591--624.

\bibitem{BrakGR04} R. Brak, J. de Gier and V. Rittenberg, 2004 {\it
  Nonequilibrium stationary states and equilibrium models with long
  range interactions}, J. Phys. A {\bf 37} 4303--4320 

\end{thebibliography}
\end{document}